# Direct observation of electronic band gap and hot carrier dynamics in GeAs semiconductor

Zailan Zhang[1#], Jiuxiang Zhang[2#], Gangqiang Zhou[3], Jiyuan Xu[3], Xiao Zhang[4], Hamid Oughaddou[4], Weiyan Qi[5], Evangelos Papalazarou[2], Luca Perfetti[5], Zhesheng Chen[3*], Azzedine Bendounan[6*] and Marino Marsi[2*]

[1]School of Physics, Nanjing University of Science and Technology, Nanjing 210094, China

[2]Laboratoire de Physique des Solides, CNRS, Université Paris Saclay, Orsay 91405, France

[3]School of Material Science and Technology, Nanjing University of Science and Technology, Nanjing 210094, China

[4]Institut des Sciences Moléculaires d'Orsay, CNRS, Université Paris-Saclay, Orsay 91405, France

[5]Laboratoire des Solides Irradiés, CEA/DRF/lRAMIS, Ecole Polytechnique, CNRS, Institut Polytechnique de Paris, F-91128 Palaiseau, France

[6]Société Civile Synchrotron SOLEIL, L'Orme des Merisiers, Départementale 128, Saint-Aubin, 91190, France

[#]These authors contribute equally to this work.
*To whom correspondence should be addressed.
zhesheng.chen@njust.edu.cn
azzedine.bendounan@synchrotron-soleil.fr
marino.marsi@universite-paris-saclay.fr

**Abstract:**

Germanium arsenide (GeAs) is a layered semiconductor with remarkably anisotropic physical, thermoelectric and optical properties, and a promising candidate for multifunctional devices based on in-plane polarization dependent response. Understanding the underlying mechanism of such devices requires the knowledge of GeAs electronic band structure and of the hot carrier dynamics in its conduction band, whose details are still unclear. In this work, we investigated the properties of occupied and photoexcited states of GeAs in energy-momentum space, by combining scanning tunneling spectroscopy (STS), angle-resolved photoemission spectroscopy (ARPES) and time-resolved ARPES. We found that, GeAs is an indirect gap semiconductor having an electronic gap of 0.8 eV, for which the conduction band minimum (CBM) is located at the $\bar{\Gamma}$ point while the valence band maximum (VBM) is out of $\bar{\Gamma}$. A Stark broadening of the valence band is observed immediately after photoexcitation, which can be attributed to the effects of the electrical field at the surface induced by inhomogeneous screening. Moreover, the hot electrons relaxation time of 1.56 ps down to the CBM which is dominated from both inter-valley and intra-valley coupling. Besides their relevance for our understanding of GeAs, these findings present general interest for the design on high performance thermoelectric and optoelectronic devices based on 2D semiconductors.

**Keywords**: Germanium arsenide; ARPES; electronic band structure; band gap; hot carrier dynamics

# INTRODUCTION

Layered in-plane anisotropic semiconductors are the subject of continuous attention, due to their anisotropic electrical and optical properties, which can be instrumental in the development of polarized electronic and optoelectronic devices such as polarization-sensitive photodetectors, linearly-polarized pulses generators and high performance thermoelectric devices[1-9]. Black phosphorus (BP) is the most representative material of this class of semiconductors, with extraordinarily anisotropic properties: the large difference between electron and hole effective mass along the armchair and zigzag symmetry directions results in a mobility ratio of ~4 for a few-layer BP along the former with respect to the latter[10-15]. In GeAs, a mobility anisotropy ratio as high as 4.6 has recently measured[1]. In addition, the hole carrier mobility is found to approach 100 cm$^2$ V$^{-1}$ s$^{-1}$ with ON–OFF ratio over 10$^5$, well comparable with state-of-the-art transition metal dichalcogenide (TMD) devices[2]. Furthermore, it exhibits superior thermoelectric properties, demonstrating a notable figure of merit (ZT) reaching a maximum value of 0.35 at 660 K[3]. GeAs excellent thermoelectric properties are attributed to a contribution of multiple large valence bands and the small differences in energy between multi-valleys around the valence band maximum (VBM)[16]. Thus, it is crucial to explore in detail its valence band dispersion.

The type and size of band gaps in semiconductors determine their optical transitions, which in turn are responsible for the optical absorption and photo-response range in optoelectronic devices. According to first-principle calculations, few-layer GeAs and bulk GeAs have indirect band gaps[17,18], meaning that the states between CBM and VBM hold different wavevectors. Photoluminescence and optical absorption measurements performed in few-layer GeAs have estimated an optical gap size of 0.5 eV-0.8 eV[19]. Up-to-date, however, there is no direct measurement of the electronic band gap size and wavevectors. Their knowledge makes it crucial to understand electronic dynamics. The photo-excited electrons in the conduction band play a key role in the performance of optoelectronic devices, which is dominated by electron-phonon coupling. Overall, our understanding of the properties of the band gap and of photoexcited states can strongly benefit from ultrafast pump-probe measurements with ultrahigh time, energy and momentum resolutions. ARPES is one of the most direct methods to investigate the electronic band structure of a solid with both high energy and momentum resolutions[20,21]. In this work, the valence band dispersion in GeAs has been systematically investigated thus revealing a VBM located at *$k_x$≈0.08 Å$^{-1}$*. However, the limitation of ARPES is that it can only observe occupied electronic states: consequently, the conduction band as well as band gap cannot be visualized by ARPES in most semiconductors because the Fermi level is located in their band gap. The recent development of time-resolved ARPES (tr-ARPES) overcomes this limitation, since the electrons can be photoexcited from the valence band to the conduction band and then they are photoemitted before recombination[22,23].

In order to gain more insights on the bandgap properties and ultrafast hot electron dynamics in the conduction band, we investigated photoexcited states in GeAs by using tr-ARPES. The band gap size is determined at 0.8 eV, which is in consistent with scanning tunneling spectroscopy (STS) results, and the indirect nature of the band gap is confirmed as the conduction band minimum (CBM) and VBM are located at different wavevectors. The relaxation process of hot electrons in $\bar{\Gamma}$ valley is dominated by long-range Fröhlich coupling. Overall, our investigations on both occupied and photoexcited states in GeAs provide critical information on the electronic band structure and hot carrier dynamics in the conduction band, essential for the design of high-performance multifunctional devices in the future.

**RESULTS AND DISCUSSION**

Bulk GeAs possesses a monoclinic crystal structure in the space group C2/m (No. 12), in which Ge forms different kinds of Ge-Ge bonds with two orientations: one is almost within the layer plane while the other one is out of the plane, as described in Fig. 1a-b. The lattice constants of GeAs are a = 3.833 Å, b = 8.453 Å, and c = 9.902 Å[5]. GeAs single crystals have been purchased from 2D Semiconductors. The anisotropic properties of GeAs can be determined by the Raman spectroscopy and angle-resolved polarized Raman yield from the vibration mode at 147 cm$^{-1}$, as shown in Fig. 1b-c. In Fig. 1d, spin-orbit splitting of As and Ge 3p core levels can be clearly detected with x-ray photoemission spectroscopy (XPS), which indicates the high quality of the sample. GeAs cleaves naturally along (111) using silver epoxy glue. Figure 1e shows the surface morphology as it was monitored by atomic-resolution STM. The armchair and zigzag directions are labeled with a white dash line and a blue line, respectively. Figure 1f illustrates the 3D Brillouin zone (BZ) of GeAs, and the surface Brillouin zone which projects the bulk BZ onto the (111) plane of the reciprocal space.

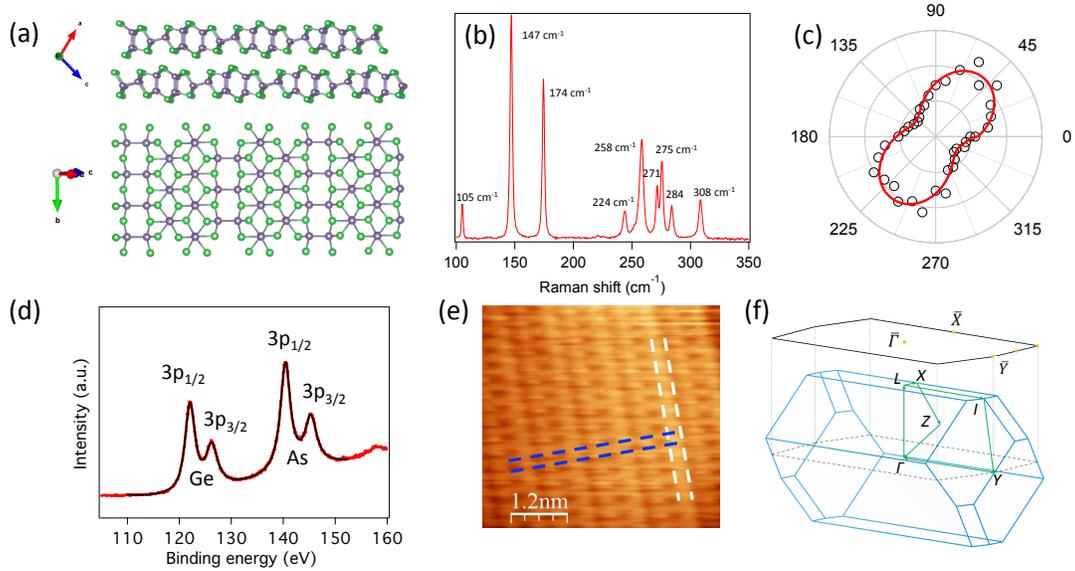

**Fig. 1: Schematics and characterizations of a GeAs single crystal**. **a** Crystal structure of the GeAs compound. **b,c** Raman shift of GeAs and polarized Raman intensity of a vibrational mode at 147 cm$^{-1}$. **d** XPS taken with a photon energy of 820 eV. The spectrum is fitted using Voigt functions (black curve). **e** Atomically resolved STM image of a cleaved GeAs(111) single crystal. **f** The bulk Brillouin zone (BZ) and surface BZ of GeAs.

The electronic band structure of GeAs is directly observed from ARPES experiments under photon energy of 70 eV as shown in Fig. 2, which we carried out at the TEMPO beamline of Synchrotron SOLEIL. From the constant energy maps shown in Fig. 2a, one can observe high symmetry directions and a rectangular feature with two-fold in-plane symmetry, indicating an anisotropy in the electronic band structure of GeAs. The VBM is located at 0.1 eV below the Fermi level, a confirmation of the p-type nature of pristine GeAs single crystals. The isoenergy cut at $E-E_F=-0.1$ eV reveals the VBM to be located at $k_x \approx 0.08$ Å$^{-1}$, instead of $\bar{\Gamma}$ point. In Fig. 2b-c, we further investigate band dispersions along two high symmetric directions, $\bar{\Gamma}-\bar{X}$ and $\bar{\Gamma}-\bar{Y}$. The sharp contrast of dispersions along $\bar{\Gamma}-\bar{X}$ and $\bar{\Gamma}-\bar{Y}$ leads to a high anisotropy of electrons in their propagation along different in-plane directions. From first principle calculations of other literatures, there is also a large band dispersion along the $k_z$ direction[18]. In ARPES experiments, $k_z$ can be extracted by photon energy $h\nu$ dependent measurements according to $k_z = \sqrt{2m(E_k cos\theta^2 + V_0)/\hbar^2}$ where $\theta$ is the emission angle and $E_k$ is the kinetic energy of the emitted free electrons satisfying $E_k = h\nu - \varphi - E_B$ where $\varphi$ is the work function of the sample and $E_B$ is the electron binding energy. The ARPES data taken from different photon energies $h\nu$ = 70 eV, 80 eV, 90 eV and 100 eV are shown in Fig. 2d. Surprisingly, the VBM does not show photon energy dependent features which is in stark contrast with the large $k_z$ dispersion predicted by calculations. The corresponding energy distribution curves (EDCs) further confirm that the VBMs is located at E-E$_F$=-0.1 eV, while its value is E-E$_F$=-0.2 eV at $\bar{\Gamma}$, for all the photon energies. We attribute this considerably weak VBM dispersion to strong bulk-surface interaction as it has been observed in another anisotropic material GaTe[24]. In addition, the spin-orbit induced shift on the first valence band along $k_z$ direction was underestimated in theoretical calculations, which can be another reason for the disagreement between calculations and experimental results[25].

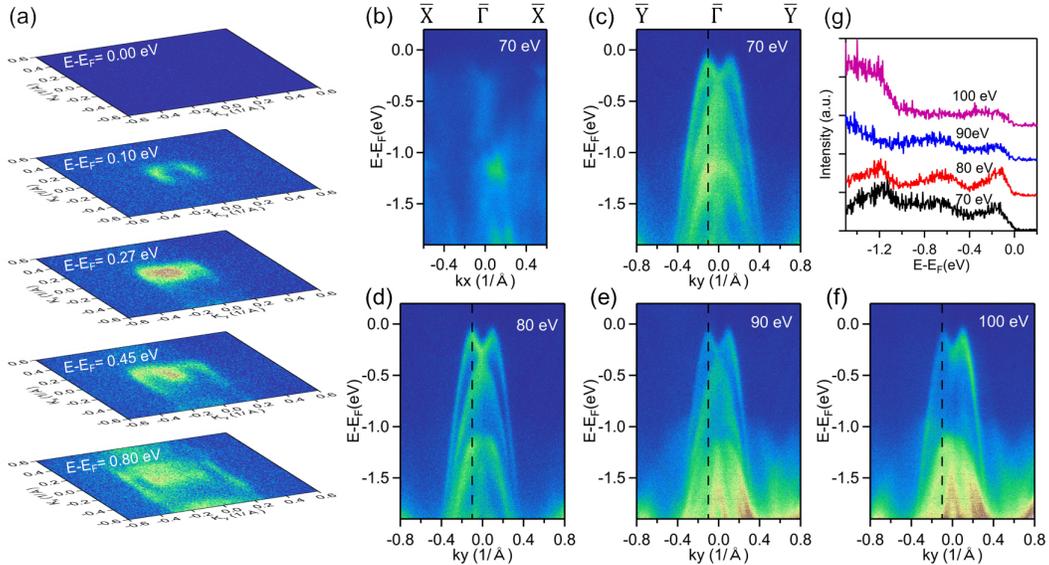

**Fig. 2: ARPES of GeAs. a** Stacking plots of constant-energy contours at different

binding energies. **b-c** Dispersions of GeAs measured along $\bar{\Gamma} - \bar{X}$ and $\bar{\Gamma} - \bar{Y}$ under photon energies of 70 eV. **d-f** Dispersions of GeAs measured along $\bar{\Gamma} - \bar{Y}$ under photon energies of 80 eV, 90 eV and 100 eV, respectively. **g** EDCs taken at the VBM from c-f.

The properties of electronic band gap of GeAs were further investigated by both STS and tr-ARPES experiments, which are shown in Fig. 3. STS can provide information about the local density of states; the band gap can be extracted from the differential conductance (dI/dV) spectrum of the sample. The STS results in Fig. 3a show that the VBM of GeAs is located at $E-E_F$=-0.1 eV and the CBM appears at $E-E_F$=0.7 eV above Fermi level which gives a bandgap of $E_G$=0.8 eV. The band gap size obtained from STS is consistent with that extracted from optical measurements and previously reported[18] theoretical calculations. However, STS measurements lack of momentum information, so that the type of band gap is still unknown. In the following, we present the results obtained from tr-ARPES experiments, which possess the ability to observe both valence band and conduction band dispersions in energy-momentum (E-k) space. Ultrashort infrared pump pulses (1.57 eV) and ultraviolet probe pulses (6.3 eV) were used to excite electrons from valence band to conduction band and subsequently photoemit them. Fig. 3b shows the transient dispersion of GeAs along $\bar{\Gamma} - \bar{X}$ and $\bar{\Gamma} - \bar{Y}$ directions at a pump-probe delay time of 500 fs. In addition to the VBM, one can see the CBM at around $E-E_F$=0.79 eV. The valence band observed from tr-ARPES shows the similar dispersion with the first valence band from synchrotron based ARPES, i.e. the VBM is out of $\bar{\Gamma}$ point. However, for the conduction band dispersion, it is evident that the CBM is at the $\bar{\Gamma}$ point, which indicates an indirect bandgap. The dispersion of the CBM can be fitted by a parabolic curve giving an effective mass of *m\*=0.18±0.03m$_e$* along $\bar{\Gamma} - \bar{X}$ direction and *m\*=0.55±0.05m$_e$* along $\bar{\Gamma} - \bar{Y}$ direction for electron, where $m_e$ is free electron mass. The corresponding EDCs along $\bar{\Gamma} - \bar{Y}$ direction at $k_x$=0 Å$^{-1}$ (for CBM) and $k_x$=0.08 Å$^{-1}$ (for VBM) are plotted in Fig. 3c. The sharp contrast of the electron's effective mass along $\bar{\Gamma} - \bar{X}$ and $\bar{\Gamma} - \bar{Y}$ directions again demonstrates the in-plane anisotropic properties in GeAs.

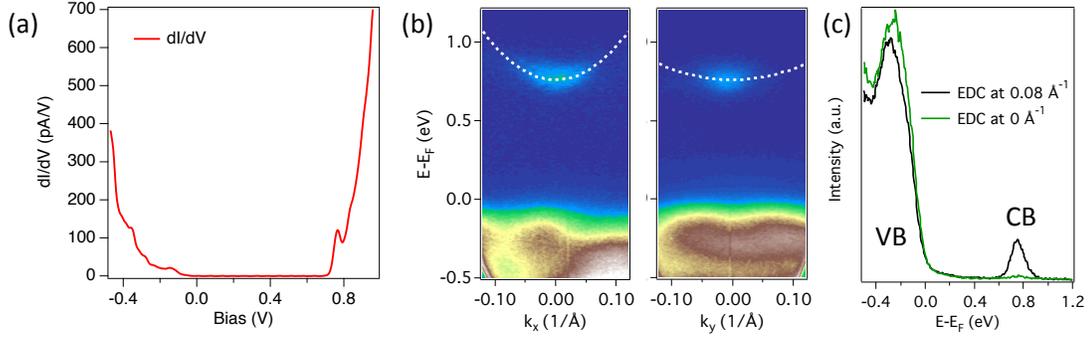

**Fig. 3: Electronic band gap of GeAs. a** dI/dV spectra of GeAs single crystal. **b** Time-resolved ARPES of GeAs along armchair and zigzag directions under pump-probe delay of 500 fs. The conduction band was fitted by white dashed curves. **c** The corresponding EDC curves of time-resolved ARPES data taken from zigzag direction at k=0 Å$^{-1}$ and 0.08 Å$^{-1}$.

Let us know discuss the hot carrier dynamics. Besides the transient tr-ARPES intensity maps acquired at a pump-probe delay of 500 fs, in Fig. 4a we plot the differential intensity maps at various delays, 0, +0.2, +0.5, +1, +2 ps, and an intensity map at delay time of -1 ps. At time delay zero, hot electrons populate high-energy states in the conduction band. Electrons dissipate energy following scattering processes and they reach the CBM. We also observe apparent extra electrons distributed around the first valence band (red band close to the Fermi level) at pump-probe delays within 1 ps: this actually indicates an energy shift of the valence band due to Stark effect that is further caused by inhomogeneous potentials along the surface plane [26]. Note that the charged impurities are present in *p*-type GeAs according to theory prediction and carrier transport measurement by J. Sun *et al*[27]. The acceptors in the near GeAs surface are negatively charged, thus generating localized states. After photoexcitation, the injected carriers scatter strongly with the localized states. Therefore, an inhomogeneous screening of the local potential lead to electrical field parallel to the surface plane. The Stark broadening of valence band after photoexcitation has also been observed in other 2D *p*-type materials, such as black phosphorus as we reported before[12].

Next, we focus on the hot electron dynamics. The dynamics of photoelectron intensity integrated in the wavevector interval [-0.1, 0.1] Å$^{-1}$ is shown in Fig. 4b. The average excess energy <E> of the electrons distributed in conduction band is obtained by $\int EI(E,t)dE / \int I(E,t)dE$ over 0.5 eV<E<1.2 eV, as shown in Fig. 4c. It can be seen from the results that the <E> has an initial value of 0.85 eV and follows an exponential decay down to 0.68 eV. A characteristic time constant τ=1.56±0.1 ps has been subsequently extracted. Such slow hot electron relaxation is attributed to both inter-valley coupling and intra-valley coupling processes, even though we can only observe the hot carrier distribution in $\bar{\Gamma}$ valley in our experiment as the other valley is far away in momentum space. The corresponding schematic is shown in Fig.4d. The intra-valley scattering in $\bar{\Gamma}$ valley is dominated by long-range Fröhlich interaction to phonons, which is universal to polar semiconductors (like for instance in InSe). From the dynamics of <E>, we can estimate the cooling rate of hot electrons λ simply by Δ<E>/ τ and obtained value λ=0.11eV/ps in GeAs is much less than the cooling rate in InSe under the same experimental conditions[28].

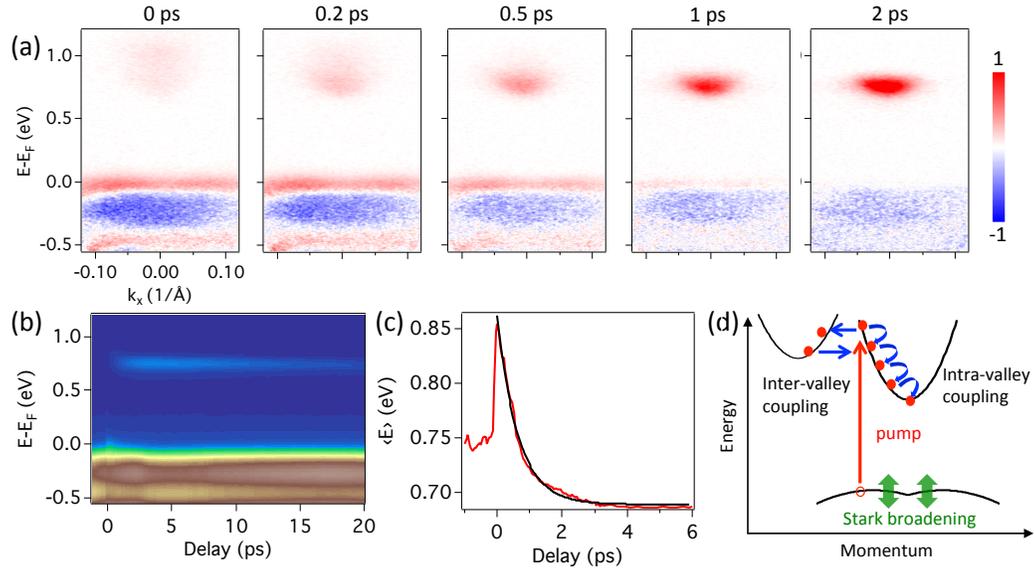

**Fig. 4: Hot carrier dynamics of GeAs. a** Difference between intensity maps at delays 0, 0.2, 0.5, 1, 2 ps and an intensity map at a delay -1ps. The data have been acquired along $\bar{\Gamma} - \bar{Y}$ direction with a pump fluence of 120 μJ/cm². The red and blue color maps represent electrons and holes, respectively. **b** Dynamics of photoelectron intensity integrated in the wavevector interval [-0.1, 0.1] Å⁻¹. **c** Average of kinetic energy of the electrons in the conduction band as a function of pump-probe delay. The solid black line

is an exponential fit with decay time τ=1.56 ps. **d** Schematic of inter-valley scattering and intra-valley scattering in GeAs.

## CONCLUSION

In conclusion, we investigated the electronic band structure and hot carrier dynamics in the anisotropic 2D semiconductor GeAs. A weak $k_z$ dependence in the valence band was observed experimentally, which might be due to strong bulk-surface interaction. An indirect bandgap of 0.8 eV was observed using tr-ARPES, while its value was extracted from both STS and tr-ARPES measurements. Broadening of valence band states observed upon photoexcitation is consistent with Stark effect, which assumes an electrical field parallel to the surface created by inhomogeneous screening of a local potential. The decay dynamics of hot electrons in the conduction band was further analyzed, obtaining a value of 1.56 ps. This decay value can be attributed to both inter-valley and intra-valley coupling of mobile carriers with phonons. Our experimental results not only provide critical information on the electronic band structure and hot carrier dynamics, but also pave a way to design efficient electronics and optoelectronic devices based on GeAs.

## METHODS

**STM and STS experiments.** Single crystal of GeAs semiconductor were purchased from 2D semiconductors company and cleaved in ultrahigh vacuum chamber at base pressure better than $5\times10^{-10}$ mbar for all the experiment. STM measurements were operated at 77 K and recorded at a constant current mode. The tungsten (W) tips were made by electrochemical etching. STS experiments were measured at 77 K using a standard lock-in technique, where a modulation voltage of 10 mV at a frequency of 5127.7 Hz was applied.

**ARPES and XPS experiments.** ARPES and XPS experiments were carried out at TEMPO beamline of Synchrotron SOLEIL, France. High quality of GeAs samples were cleaved at the base pressure of $1\times10^{-10}$ mbar under temperature of ~80 K. The photoelectrons were collected with new generation MBS A-1 analyzer equipped with electrostatic lens allowing Fermi surface mapping in order to determine the sample orientation. The energy and momentum resolutions were better than 35 meV and 0.01 Å$^{-1}$.

**Time-resolved ARPES experiments.** Tr-ARPES experiments were performed on the femto-ARPES platform using a Ti/sapphire laser system, delivering 50 fs pulses at 1.57 eV (790 nm) with a 250 kHz repetition rate[29-31]. Part of the laser beam is used to generate 6.3 eV photons through cascade frequency mixing in $BaB_2O_4$ (BBO) crystals. The 1.57 and 6.3 eV beams were employed to photoexcite the sample and induce photoemission. The overall energy resolution of the experiment is about 30 meV, whereas the cross-correlation between pump and probe pulse has a full width at half-maximum (FWHM) of 0.16 ps. The samples were kept at ~120 K during measurements.

**Polarized Raman experiments.** Polarized Raman measurements were performed on Horiba Raman spectroscopy. A continuous laser with a wavelength 532 nm through 100X objective lens were used to excite the sample. The incident light passing through a linear polarizer was reflected by a beam splitter. A half-wave plate was positioned between the beam splitter and the objective lens in order to rotate the polarization of the incident light. Another half-wave plate was positioned in the collection path, followed by a linear polarizer to select the polarization of interest from the scattered light, which was subsequently detected by a CCD camera. The samples were rotated at a step of 10° during measurements.

## ACKNOWLEDGEMENTS

We acknowledge financial support from "Investissement d'Avenir" Labex PALM (ANR-10-LABX-0039-PALM) and from Nanjing University of Science and Technology (AE899991/406, AD411203). We thank synchrotron SOLEIL for provision of ARPES beamtime under proposal of No. 20220536. J. Zhang thanks the financial support of China Scholarship Council.